\documentclass[aps,prl,twocolumn,superscriptaddress,floatfix,10pt]{revtex4-1}
\usepackage{graphicx}% Include figure files
\usepackage{dcolumn}% Align table columns on decimal point
\usepackage{bm}% bold math
\usepackage{amsfonts}
\usepackage{amssymb}
\usepackage{amsmath}

\usepackage{bbold}

\newcommand{\oeq}{\begin{equation}}
\newcommand{\ceq}{\end{equation}}
\newcommand{\oeqn}{\begin{eqnarray}}
\newcommand{\ceqn}{\end{eqnarray}}

\renewcommand{\>}{\rangle}
\newcommand{\<}{\langle}

\newcommand{\oQ}{\hat{Q}}

\newcommand{\oR}{\hat{R}}

\begin{document}

\title{Formation and dynamics of fission fragments}

\author{C. Simenel}
\affiliation{Department of Nuclear Physics, Research School of Physics and Engineering, Australian National University, Canberra, Australian Capital Territory 0200, Australia}

\author{A. S. Umar}
\affiliation{Department of Physics and Astronomy, Vanderbilt University, Nashville, Tennessee 37235, USA}

\date{\today}

\begin{abstract}
Although the overall time-scale for nuclear fission is long, suggesting a slow process, rapid shape evolution occurs in its later stages near scission.
Theoretical prediction of the fission
fragments and their characteristics are often based on the assumption that the internal degrees of freedom are equilibrated along the fission path.  However, this adiabatic approximation may break down near scission.  
 This is studied for the symmetric fission
of $^{258,264}$Fm. The non-adiabatic
evolution is computed using the time-dependent Hartree-Fock method, starting from an adiabatic configuration where the fragments have acquired their identity. It is shown that dynamics has an important
effect %R1-1 on the scission configuration and 
on the kinetic and excitation energies of the fragments. The vibrational modes
of the fragments in the post-scission evolution are also analyzed.
\end{abstract}
%\keywords{}
\maketitle

%The competition between the attractive strong interaction and Coulomb repulsion in heavy atomic nuclei can induce an instability against separation into two lighter fragments. 
%Since its discovery in 1938~\cite{hah39,mei39}, the nuclear fission process has been widely studied~\cite{wag91}. 
Fission can be found in different complex quantum systems, such as atomic nuclei ~\cite{hah39,mei39} and atomic clusters \cite{fra01}.
This is  one of the most challenging quantum many-body problems,   
%Yet, a fully microscopic description of nuclear fission remains as one of the most elusive problems in quantum many-body physics.  
due to the difficulty of finding an adequate and computationally tractable formulation of the evolution from the compound system to the formation of the final fragments.
For atomic nuclei, the minimum average time scale for such an evolution is of the order of $20-50$~zs \cite{hin92}.
%The complexity of the fission dynamics could be partially illustrated by the large variations in fission times,  reaching $10^{-16}$~s in the case of statistical fission of compound nuclei formed in fusion reactions~\cite{and07}, and even much longer times if fission occurs by quantum tunneling through the fission barrier. 
%These times are extremely slow in comparison to the typical time, $t\sim10^{-22}$~s, for a nucleon to traverse the nucleus. 
 This is slow enough to consider, as a first approximation, nuclear fission as an adiabatic process. 
 This means that the nucleonic degrees of freedom are fully equilibrated while the system evolves over a potential energy surface (PES) defined by the macroscopic degrees of freedom such as elongation and mass asymmetry~\cite{mor91}.

However, the adiabatic approximation is expected to break down in the final stages of the fission process, when scission of the neck between the fragments occurs \cite{riz13}. In this phase, fragments can experience a rapid change in shape~\cite{dub08},
which
would be better described with a non-adiabatic approach. A realistic description of the entire fission process could
then be achieved with an adiabatic model describing the slow evolution across the barrier, followed by a non-adiabatic
treatment of the scission and post-scission dynamics. The transition between the adiabatic and non-adiabatic pictures is
expected to occur somewhere between the top of the fission barrier and  the scission point. It is desirable that these
two methods are based on a consistent approach to the many-body problem.

To date, most of the theoretical works have focused on the adiabatic part of the fission process. 
%, assuming that
%the characteristics of the fragments (mass, charge, kinetic energy...) could be estimated from the configuration of the
%system at the (adiabatic) scission point~\cite{gou05,bon06,ran11}. %Amongst these models, one of the most advanced
%description of fission has been provided by the time-dependent generator coordinate method (TDGCM) with the Gaussian
%overlap approximation~\cite{gou05}. %In this approach, an adiabatic PES is generated from a set of mean-field
%calculations with constraints on few most relevant macroscopic degrees of freedom. %The collective wave function is then
%evolved in the PES. %Formal attempts to include adiabatic effects on top of this approach by including couplings between
%collective and intrinsic degrees of freedom have recently been made~\cite{ber11}. %This approach being computationally
%demanding, it has been applied to induced fission of $^{238}$U only. 
Microscopic approaches have been widely used to study  fission paths (see Refs.~\cite{gou05,bon06,dub08,sta09,pei09,you11,war12,abu12,mir12,sta13,mcd13,sad13,sch14a,sch14b} for recent applications).
In particular, the time-dependent generator coordinate method~\cite{gou05} and the adiabatic time-dependent Hartree-Fock theory~\cite{bar78} provide a description of the evolution of collective and internal degrees of freedom. %~\cite{sta13}. 
Simplifications using the strongly damped character of fission have also been widely used. 
% to avoid the numerical difficulties inherent to the above (semi-)microscopic approaches. 
For instance, random walks on a phenomenological five-dimensional
PES~\cite{mol01}, in analogy with Brownian motion, have led to a good description of fragment mass distributions
\cite{ran11}.

All of these approaches  aim to describe the fission process up to the scission point and to predict
the properties of the fragments (mass, charge, kinetic energy), which are the main experimental observables. 
These properties are  estimated in a sudden approximation from the scission configuration assuming sharp cuts across the neck~\cite{wil76} and ignoring non-adiabatic effects coming from couplings between collective and intrinsic degrees of freedom~\cite{ber11,riz13}. 
This approximation induces a strong limitation in the predicting power of fragment characteristics.
For instance, part of the kinetic energy could come from
pre-scission dynamics~\cite{you11}. In addition, the approach is not able to describe post-scission dynamics of the fragments,
such as their vibrational modes. 
%The later stage of fission is then usually treated in a simple sudden approximation.
For this reason, it is highly desirable to go beyond this approximation and describe the later stages of the fission 
process in a dynamical and non-adiabatic fashion \cite{uma10c}.
 The time-dependent Hartree-Fock (TDHF) theory \cite{dir30} is an ideal tool to study the 
latter stage of fission as it is a fully microscopic and non-adiabatic approach. 
%Furthermore, the new time-dependent quantum microscopic approach should remain faithful to the approximations used in the 
%adiabatic stage of the fission process, using the same nucleon-nucleon effective interaction as in the adiabatic 
%calculations~\cite{uma10c}, without introducing any new parameters.
An early
attempt to describe fission with such a model was proposed in Ref.~\cite{neg78}. 
%The dynamics of scission was described with a time-dependent mean-field approach. 
Due to computational limitations, these calculations were essentially
qualitative, assuming spatial symmetries and using a simplified effective interaction. 
% without spin-orbit terms, while
A pairing gap with arbitrarily large values was also used as a phenomenological parameter to trigger scission.

Here, we investigate the formation and dynamics of fission fragments 
% of $^{264}$Fm
using a realistic three-dimensional
mean-field description. The adiabatic phase is described in the traditional way, using a static mean-field approach with an
external constraint inducing deformation. The shell structure and level crossings are used to determine at 
which deformation the
fragments have established their identity, which occurs between the saddle and scission points~\cite{mol00}. This 
determines
the initial condition for the time-dependent calculations of the non-adiabatic evolution, including scission and
post-scission dynamics.
\begin{figure}[!htb]
\includegraphics*[width=8cm]{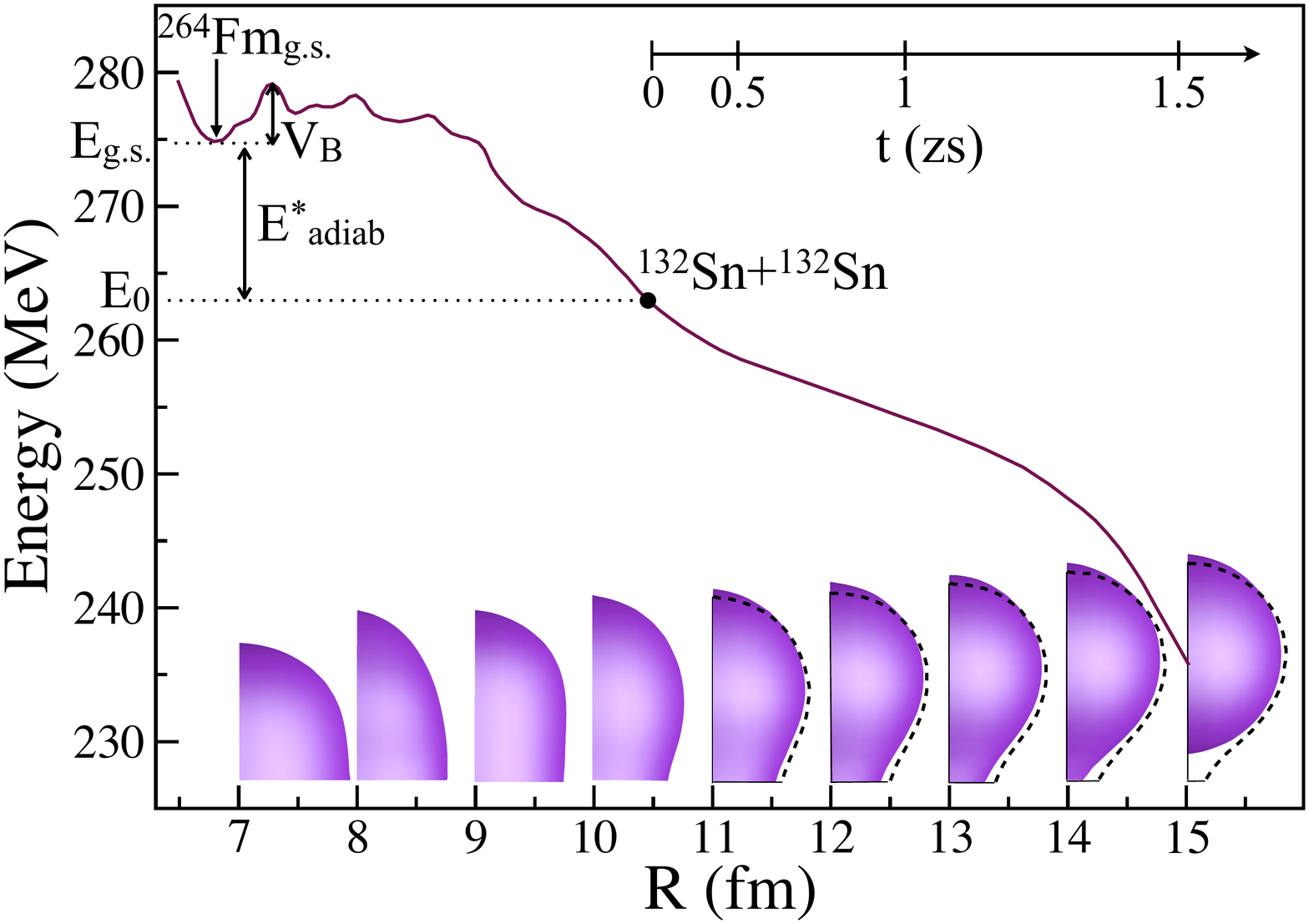}
\caption{(Color online) Adiabatic fission potential of $^{264}$Fm$\rightarrow^{132}$Sn$+^{132}$Sn (solid line) as
function of the distance $R$ between the fragment centers of mass. The reference energy is chosen such that $E=0$ for $R\rightarrow\infty$.
Adiabatic (purple surfaces) isodensities at half the
saturation density $\rho_0/2=0.08$~fm$^{-3}$ are shown at $R=7, 8\cdots 15$~fm. Half a fragment is represented, the
fission axis being vertical. TDHF isodensities are represented by dashed lines at $R=11,12\cdots15$~fm. The non-linear
axis (top) relates the time $t$ with $R$ during the non-adiabatic evolution, with origin $t=0$ associated to preformation of $^{132}$Sn fragments.}
\label{fig:potential}
\end{figure}

As a proof of concept and to establish the feasibility of this approach we studied the symmetric fission of $^{264}$Fm. 
This exotic nucleus represents an
important milestone in fission studies as it is predicted to spontaneously fission into two doubly magic $^{132}$Sn
fragments~\cite{ghe96,asa04,sad13}. 
% with a life-time of the order of $1~\mu$s~\cite{sta13}. 
Its study could be envisaged
with upcoming radioactive beams or, alternatively, using multi-nucleon transfer reactions in actinide collisions, which
have been the focus of recent experimental~\cite{dvo11} and theoretical~\cite{zag06,gol09,ked10} efforts.

The adiabatic configurations are obtained by solving the Hartree-Fock (HF) equations with the BCS pairing residual
interaction (HF+BCS) using the \textsc{ev8} code~\cite{bon05}. The calculations are performed on a Cartesian grid with
mesh size $0.8$~fm. The mean-field is obtained from the SLy4$d$~\cite{kim97} Skyrme energy density functional
\cite{sky56}, and a surface pairing interaction~\cite{bender03} is used to describe the nuclear superfluid phase.
Elongations along the $z$-axis are induced by adding an external constraint $\lambda(\<\oR\>-R)z$ to the
single-particle potential. The Lagrange parameter $\lambda$ quantifies the strength of the constraint and $R$ is the
desired expectation value of
the operator $\oR$ measuring the distance between the centers of mass of the matter on each side of the neck plane 
assuming a sharp cut. %R1-2
For symmetric fission, the neck is at $z= 0$.

The adiabatic potential obtained from the constraint calculations is shown in Fig.~\ref{fig:potential}. 
% for $6.5$~fm$\le R\le15$~fm with step $\Delta R=0.1$~fm.
The fission barrier height is $V_B\simeq4.3$~MeV at $R_B\simeq7.3$~fm. 
This height is in excellent agreement with recent theoretical calculations \cite{mol09,pei09,sta09,kow10,sad13}. 
A neck is observed up to $R\simeq14.5$~fm.
%R1-1(we show later that the neck breaks at larger distances when non-adiabatic effects are included). 
It is interesting to note that the pre-scission configuration
consists of two quasi-spherical fragments. In fact, three different fission valleys have been predicted in fermium
isotopes~\cite{asa04,bon06}: {\it(i)} a mass asymmetric one, {\it(ii)} a mass symmetric one with elongated fragments,
and {\it(iii)} a mass symmetric one with compact fragments like the one in Fig.~\ref{fig:potential}. The latter is the
dominant fission path in neutron-rich fermium isotopes due to the spherical shells in the vicinity of $^{132}$Sn
\cite{hul86,ich09}.

\begin{figure}[!htb]
\includegraphics*[width=8cm]{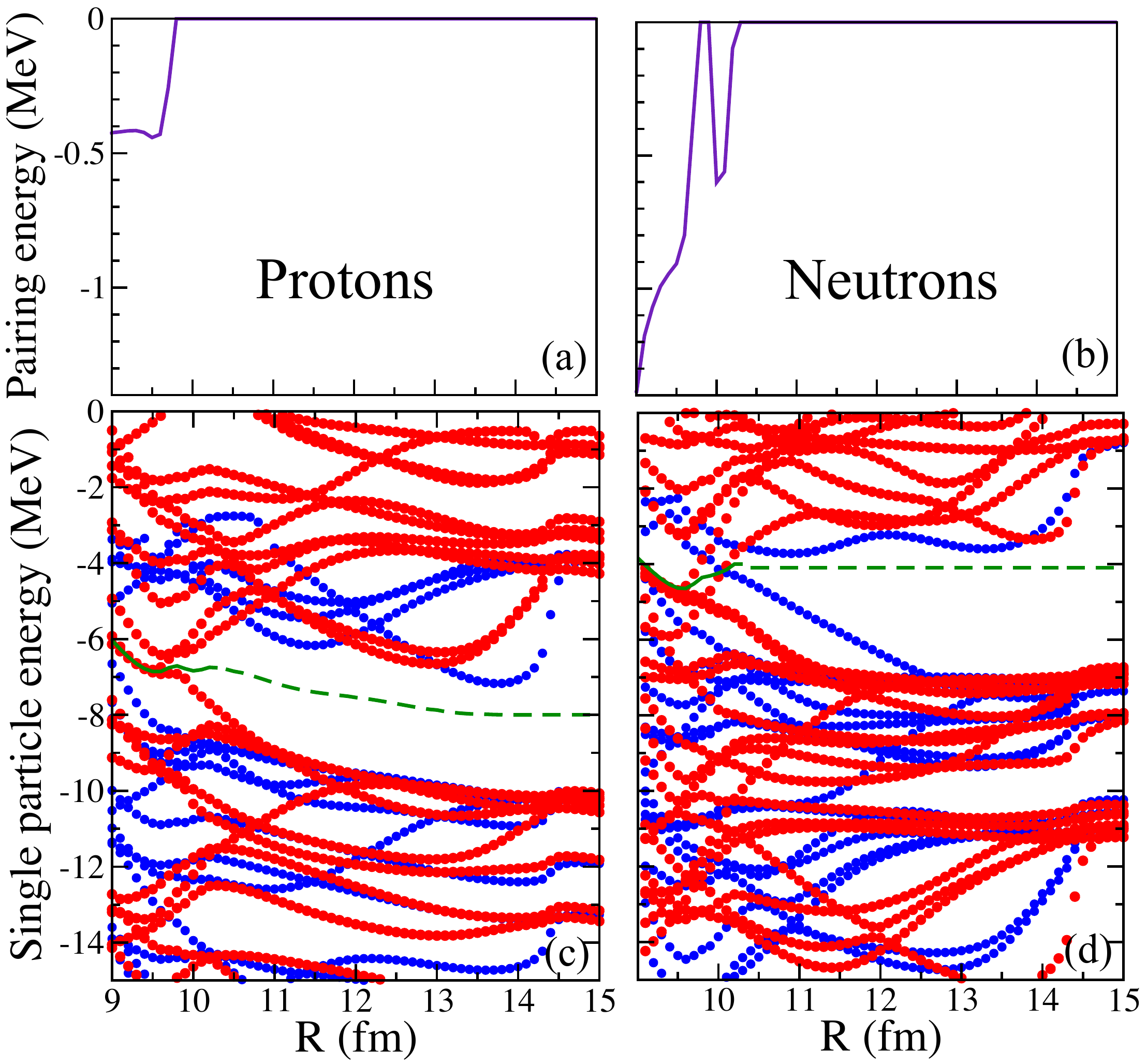}
\caption{(Color online) Proton (a) and neutron (b) pairing energy as function of $R$. Proton (c) and Neutron (d) single-particle energies for states with
positive (blue) and negative (red) parity. The green solid lines indicate the Fermi level in the presence of pairing. These are continued by 
green dashed lines, which represent the Fermi level located arbitrarily in the magic gaps.}
\label{fig:pairing}
\end{figure}

The transition criteria between adiabatic and non-adiabatic phases has yet to be defined. 
%A natural choice is to presume that
%the evolution becomes non-adiabatic when the fragments have established their identity. 
%Although it may be difficult to determine from the total density 
%The 
A distance $R_0$ where fragments are pre-formed can be determined by examining the shell structure of the system. 
The proton and neutron single-particle energies are plotted near the Fermi level for $R>9$~fm in Figs.~\ref{fig:pairing}(c) and~\ref{fig:pairing}(d), respectively. We observe that the shell gaps,
$Z=50$ and $N=82$, associated with $^{132}$Sn appear after $R\approx10$~fm. This is also confirmed by the evolution of the proton and neutron pairing
energies shown in Figs.~\ref{fig:pairing}(a) and~\ref{fig:pairing}(b), respectively.
The latter vanish around the same point as the pairing residual
interaction is not able to scatter Cooper pairs across the magic shell gaps \cite{rin80}. 
We consider that the fragments are
pre-formed at this point and experience a non-adiabatic evolution from $R_0\simeq10.5$~fm onwards, 
corresponding to the vanishing of all pairing energies.

Consistent with the adiabatic phase, the non-adiabatic evolution is computed at the mean-field level with the TDHF theory~\cite{dir30}. The latter has been widely used to investigate low-energy
nuclear dynamics (see Refs.~\cite{neg82,sim12b,sim13a} for reviews). 
Although one-body dissipation mechanisms are well accounted for in the TDHF approach, 
it does not include the Landau-Zener effect which is crucial to properly describe dissipation 
when single-particle levels with different occupation numbers cross. 
This effect could be partly accounted for with the inclusion of pairing correlations \cite{blo76}
which have been the subject of several recent works \cite{ave08,eba10,ste11,sca13}.
Here, the transition between adiabatic and non-adiabatic regimes is supposed to occur after the last crossing. 
Consequently, the Landau-Zener effect is not expected to affect the dynamics in the non-adiabatic phase.

The \textsc{tdhf3d} code~\cite{kim97} is used with
a mesh spacing of 0.8~fm and a time step $1.5\times10^{-24}$~s. 
%the same mesh spacing as in the HF+BCS calculations. 
The $z=0$ plane represents the plane of symmetry. The Cartesian grid extends to $16$~fm from the center in $x$ and $y$
and to $64$~fm in $z$ direction. 
Pairing is not included in the dynamics as the fragments maintain their double-magicity at all times. 
%The TDHF equation of motion is solved iteratively in time with a time-step $\Delta t=1.5\times10^{-24}$s.

%R1-1 Adiabatic and 
TDHF isodensities are shown in Fig.~\ref{fig:potential}. 
In this case, the neck remains at elongations up to $R\simeq15.4$~fm.
%R1-1 in TDHF due to non-adiabatic effects. 
It is interesting to quantify the time needed
for the non-adiabatic evolution to reach scission. The axis shown in the top of  Fig.~\ref{fig:potential} indicates at
which times different values of $R$ are reached. Scission occurs after $\sim1.6$~zs of non-adiabatic evolution. This
time is too short for the system to find the minimum of the
potential energy surface around scission, which is why the scission point is found to be different when non-adiabatic
effects are included.
\begin{figure}[!htb]
\includegraphics*[width=8cm]{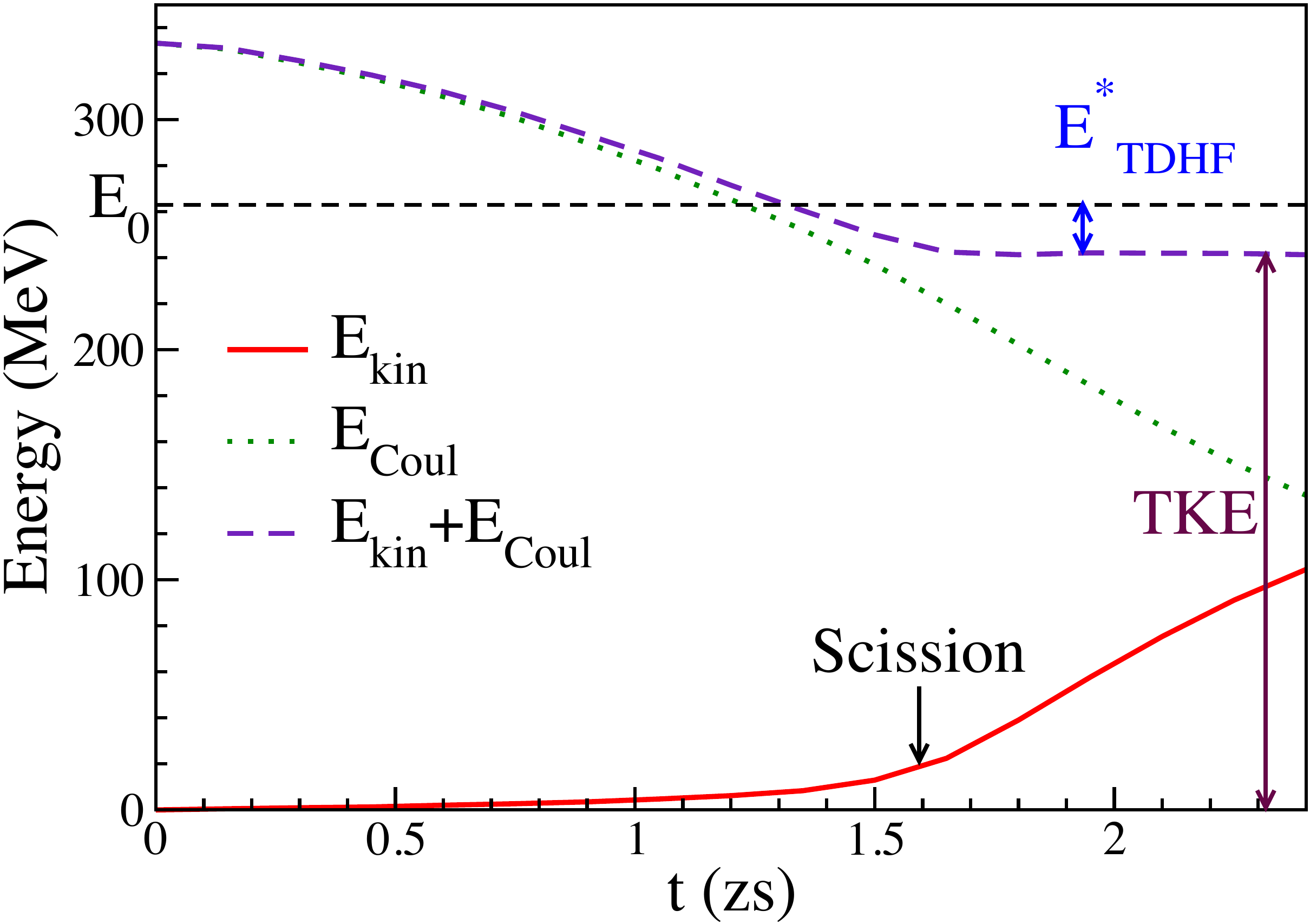}
\caption{(Color online) Time evolution of various energies in the non-adiabatic phase (see text).}
\label{fig:energy}
\end{figure}

\begin{figure*}[!htb]
\includegraphics[width=14cm]{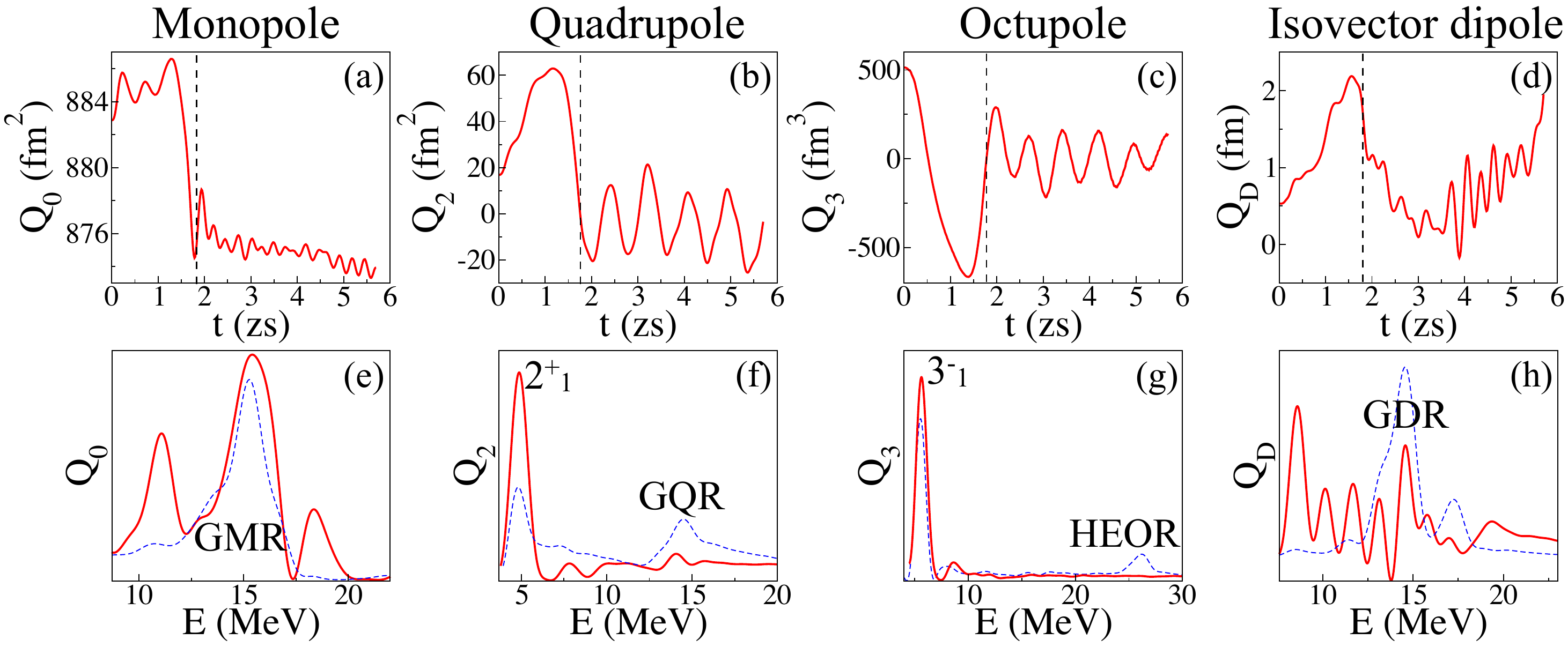}
\caption{(Color online) (a-d) Evolution of multipole moments in the non-adiabatic phase. (e-h) Fourier transforms are computed for $t>t_1=1.8$~zs (right of dashed line in top panels). The RPA strength functions of $^{132}$Sn are plotted with dashed lines (bottom) 
%obtained from small amplitude excitations on the $^{132}$Sn ground state 
 with arbitrary normalization. }
\label{fig:moments}
\end{figure*}

We now investigate the effect of the non-adiabatic evolution on the  kinetic energy $E_{kin}$ of the fragments.
Fig.~\ref{fig:energy} shows the evolution of $E_{kin}$ and of the potential energy $E_{Coul}$ arising from mutual Coulomb repulsion. 
Note that, before the fragments are well separated, these energies could depend on the definition of the fragments and on localization effects \cite{you11}. %R1-2
When the neck breaks at $t\simeq1.6$~zs, the fragments have already acquired a kinetic energy
$E_{kin}\simeq19$~MeV. This non-adiabatic contribution to the kinetic energy is usually neglected in models based on the
adiabatic approximation~\cite{you11}. The total kinetic energy (TKE) corresponding to the asymptotic value of $E_{kin}$
can be obtained by summing the Coulomb and kinetic energies when the nuclear attraction between the fragments vanishes.
We get from Fig.~\ref{fig:energy} a TKE of $\sim241$~MeV. This TKE is much larger than the prediction from the Viola
systematics~\cite{vio85} which is $\sim192$~MeV. 
This effect, already observed in lighter fermium isotopes \cite{hul86}, can be attributed to the strong spherical shell effects in the fragments which are responsible for the compact shape at scission~\cite{ich09}.

A similar analysis can be performed for the excitation energy of the fragments. If we consider spontaneous fission,
the $^{264}$Fm is initially in its ground state, corresponding to the first potential well at $R\simeq6.8$~fm with an energy $E_{g.s.}\simeq275$~MeV in Fig.~\ref{fig:potential}. At $R_0\simeq10.5$~fm, where the transition between adiabatic and
non-adiabatic regimes occurs, the potential is at $E_0\simeq263$~MeV and the system has acquired an excitation energy
during its adiabatic evolution $E^*_{adiab}=E_{g.s.}-E_0\simeq12$~MeV. 
Note that $E^*_{adiab}$ is much larger than the pairing energy and our conclusions are not affected by the choice of the pairing strength. 
During the non-adiabatic evolution, and up to
scission, the excitation energy keeps increasing due to dissipation mechanisms. The TDHF approach incorporates the
one-body dissipation mechanisms which are dominant at low energy~\cite{bon76,koo77}. As a result, the TDHF prediction of the
excitation energy acquired during the non-adiabatic phase is $E^*_{TDHF}=E_0-$TKE$\simeq22$~MeV. The asymptotic value of
the total excitation energy is then $E^*=E^*_{adiab}+E^*_{TDHF}\simeq34$~MeV. The non-adiabatic evolution is then
responsible for almost $\sim2/3$ of the final excitation energy of the fragments.

Further inquiry is required to get a deeper insight concerning the nature of the excitation energy acquired during
the non-adiabatic evolution. In the present case, the magic gap of $^{132}$Sn is expected to hinder incoherent particle-hole
excitations and subsequent thermalization of the fragments. However, collective vibrations, some of which lie at low
excitation energies, could be easily excited. Such collective modes are accounted for in the TDHF framework
\cite{bon76,blo79}. The TDHF simulation of the post-scission evolution of the fragments can then be used to investigate the
excitation of such vibrations.

Figures~\ref{fig:moments}(a-d) show the evolution of different multipole moments computed for $z>0$ up to a time $T=5.7$~zs before the fragments reach the edge of the grid. At $t\ge t_1\simeq1.8$~zs, i.e., after the neck has fully disappeared, these moments exhibit oscillations associated with vibrational modes of the outgoing fission fragments. 
The Fourier transform of $[Q(t)-Q(t_1)]f(t-t_1)$, 
%where these vibrations is estimated using $\tilde{Q}(E)=\int_{t_1}^T dt [Q(t)-Q(t_1)]
%\sin[E(t-t_1)/\hb]f(t-t_1)$. Here,  and
where $t\ge t_1$ and $f(t)=\cos^2[\pi t/2(T-t_1)]$ is a filtering function to avoid spurious oscillations in the Fourier analysis, are
shown in Figs.~\ref{fig:moments}(e-h) (solid lines). They are compared with the same quantities computed
after a boost $e^{-ik\oQ}$ applied to the $^{132}$Sn HF ground-state with a boost velocity, $k$, small enough to be in the
linear regime, i.e., to have $Q(t)\propto k$ (dashed lines). This provides a numerical estimate of the RPA strength
function of $\oQ$~\cite{rin80}.

The higher energy peaks in the strength functions are associated with giant resonances. Apart from the high-energy octupole
resonance (HEOR), all giant resonances are excited in the fission fragments. However, the excitation of low-lying
collective vibrations is predominant for the octupole ($3^-_1$ state) and quadrupole ($2^+_1$ state) modes. Such
vibrations are often excited in fusion reactions \cite{das98}, as shown in recent TDHF calculations~\cite{sim13b,sim13c}. 
%(There is no low-lying monopole and isovector dipole modes in $^{132}$Sn.)
The isovector dipole response and, to a lesser extent, the monopole one, also exhibit other high energy modes not visible in
the RPA strength functions. A possible explanation is that these vibrations are built on top of a static polarization
induced by the Coulomb interaction with the other fragment. Indeed, the isovector dipole moment, which is proportional
to the distance between the proton and neutron centers of mass, is almost always positive (see 
Fig.~\ref{fig:moments}(d)). Note that couplings between collective modes in large amplitude motion~\cite{sim01} could also
induce non-linearities in the vibrational spectra~\cite{sim03,sim09}.

Finally, to test this approach with experimental data, similar calculations have been performed for $^{258}$Fm. 
In this case, neutron pairing does not vanish. 
The TDHF calculations are performed with frozen occupation numbers starting at $R_0=12-13$ fm, for which the occupation numbers are close to the post-scission ones. 
The calculated TKE are in the range 238-241 MeV, in relatively good agreement with the high-energy mode in $^{258}$Fm (TKE$\sim232$~MeV) \cite{hul86}.

Symmetric fission of $^{258,264}$Fm has been studied.
For the first time, adiabatic and non-adiabatic phases of fission are described with realistic mean-field codes.
The evolution is assumed to be
adiabatic until the fragment's identity can be established from their shell structure. Non-adiabatic effects are then
investigated employing the TDHF evolution toward scission. This non-adiabatic evolution affects 
%R1-1 the configuration at scission, as well as 
the kinetic and excitation energies of the fragments. The post-scission TDHF evolution of the
fragments is also used to analyze their vibrational modes. 
As in the case of fusion, the low-lying collective vibrations are more
easily excited than giant resonances.
The present techniques could be easily extended to other systems, including asymmetric fission, and other observables, such as neutron emission \cite{hin92}. Recent mean-field codes
\cite{mar14,sca13,eba10,ste11} including pairing could be used.
Extensions of these codes to compute mass and charge distributions of the fragments could also be used
\cite{sim10b,sim11} in order to compare with experimental data~\cite{sch00}.

%\section{acknowledgements}

P. Quentin, L. Bonneau, D. J. Hinde, M. Dasgupta, and E. Williams are thanked for useful comments and discussions. This work has been supported by the
Australian Research Council Grants No. FT120100760 and by the U.S. Department of Energy Grant No.
DE-FG02-96ER40975 with Vanderbilt University.
%, Laureate Fellowship FL110100098 and Discovery grant DP1094947. 
The calculations have been performed on the NCI National Facility in Canberra, Australia, which is supported
by the Australian Commonwealth Government.

\bibliography{biblio}

\begin{thebibliography}{66}%
\makeatletter
\providecommand \@ifxundefined [1]{%
 \@ifx{#1\undefined}
}%
\providecommand \@ifnum [1]{%
 \ifnum #1\expandafter \@firstoftwo
 \else \expandafter \@secondoftwo
 \fi
}%
\providecommand \@ifx [1]{%
 \ifx #1\expandafter \@firstoftwo
 \else \expandafter \@secondoftwo
 \fi
}%
\providecommand \natexlab [1]{#1}%
\providecommand \enquote  [1]{``#1''}%
\providecommand \bibnamefont  [1]{#1}%
\providecommand \bibfnamefont [1]{#1}%
\providecommand \citenamefont [1]{#1}%
\providecommand \href@noop [0]{\@secondoftwo}%
\providecommand \href [0]{\begingroup \@sanitize@url \@href}%
\providecommand \@href[1]{\@@startlink{#1}\@@href}%
\providecommand \@@href[1]{\endgroup#1\@@endlink}%
\providecommand \@sanitize@url [0]{\catcode `\\12\catcode `\$12\catcode
  `\&12\catcode `\#12\catcode `\^12\catcode `\_12\catcode `\%12\relax}%
\providecommand \@@startlink[1]{}%
\providecommand \@@endlink[0]{}%
\providecommand \url  [0]{\begingroup\@sanitize@url \@url }%
\providecommand \@url [1]{\endgroup\@href {#1}{\urlprefix }}%
\providecommand \urlprefix  [0]{URL }%
\providecommand \Eprint [0]{\href }%
\providecommand \doibase [0]{http://dx.doi.org/}%
\providecommand \selectlanguage [0]{\@gobble}%
\providecommand \bibinfo  [0]{\@secondoftwo}%
\providecommand \bibfield  [0]{\@secondoftwo}%
\providecommand \translation [1]{[#1]}%
\providecommand \BibitemOpen [0]{}%
\providecommand \bibitemStop [0]{}%
\providecommand \bibitemNoStop [0]{.\EOS\space}%
\providecommand \EOS [0]{\spacefactor3000\relax}%
\providecommand \BibitemShut  [1]{\csname bibitem#1\endcsname}%
\let\auto@bib@innerbib\@empty
%</preamble>
\bibitem [{\citenamefont {Hahn}\ and\ \citenamefont
  {Strassmann}(1939)}]{hah39}%
  \BibitemOpen
  \bibfield  {author} {\bibinfo {author} {\bibfnamefont {O.}~\bibnamefont
  {Hahn}}\ and\ \bibinfo {author} {\bibfnamefont {F.}~\bibnamefont
  {Strassmann}},\ }\href {\doibase 10.1007/BF01488241} {\bibfield  {journal}
  {\bibinfo  {journal} {Naturwissenschaften}\ }\textbf {\bibinfo {volume}
  {27}},\ \bibinfo {pages} {11} (\bibinfo {year} {1939})}\BibitemShut {NoStop}%
\bibitem [{\citenamefont {Meitner}\ and\ \citenamefont {Frisch}(1939)}]{mei39}%
  \BibitemOpen
  \bibfield  {author} {\bibinfo {author} {\bibfnamefont {L.}~\bibnamefont
  {Meitner}}\ and\ \bibinfo {author} {\bibfnamefont {O.~R.}\ \bibnamefont
  {Frisch}},\ }\href {\doibase 10.1038/143239a0} {\bibfield  {journal}
  {\bibinfo  {journal} {Nature (London)}\ }\textbf {\bibinfo {volume} {143}},\
  \bibinfo {pages} {239} (\bibinfo {year} {1939})}\BibitemShut {NoStop}%
\bibitem [{\citenamefont {Frauendorf}\ and\ \citenamefont
  {Guet}(2001)}]{fra01}%
  \BibitemOpen
  \bibfield  {author} {\bibinfo {author} {\bibfnamefont {S.~G.}\ \bibnamefont
  {Frauendorf}}\ and\ \bibinfo {author} {\bibfnamefont {C.}~\bibnamefont
  {Guet}},\ }\href {\doibase 10.1146/annurev.nucl.51.101701.132354} {\bibfield
  {journal} {\bibinfo  {journal} {Ann. Rev. Nucl. Part. Sci.}\ }\textbf
  {\bibinfo {volume} {51}},\ \bibinfo {pages} {219} (\bibinfo {year}
  {2001})}\BibitemShut {NoStop}%
\bibitem [{\citenamefont {Hinde}\ \emph {et~al.}(1992)\citenamefont {Hinde},
  \citenamefont {Hilscher}, \citenamefont {Rossner}, \citenamefont {Gebauer},
  \citenamefont {Lehmann},\ and\ \citenamefont {Wilpert}}]{hin92}%
  \BibitemOpen
  \bibfield  {author} {\bibinfo {author} {\bibfnamefont {D.~J.}\ \bibnamefont
  {Hinde}}, \bibinfo {author} {\bibfnamefont {D.}~\bibnamefont {Hilscher}},
  \bibinfo {author} {\bibfnamefont {H.}~\bibnamefont {Rossner}}, \bibinfo
  {author} {\bibfnamefont {B.}~\bibnamefont {Gebauer}}, \bibinfo {author}
  {\bibfnamefont {M.}~\bibnamefont {Lehmann}}, \ and\ \bibinfo {author}
  {\bibfnamefont {M.}~\bibnamefont {Wilpert}},\ }\href {\doibase
  10.1103/PhysRevC.45.1229} {\bibfield  {journal} {\bibinfo  {journal} {Phys.
  Rev. C}\ }\textbf {\bibinfo {volume} {45}},\ \bibinfo {pages} {1229}
  (\bibinfo {year} {1992})}\BibitemShut {NoStop}%
\bibitem [{\citenamefont {Moreau}\ and\ \citenamefont {Heyde}(1991)}]{mor91}%
  \BibitemOpen
  \bibfield  {author} {\bibinfo {author} {\bibfnamefont {J.}~\bibnamefont
  {Moreau}}\ and\ \bibinfo {author} {\bibfnamefont {K.}~\bibnamefont {Heyde}},\
  }in\ \href@noop {} {\emph {\bibinfo {booktitle} {The Nuclear Fission
  Process}}},\ \bibinfo {editor} {edited by\ \bibinfo {editor} {\bibfnamefont
  {C.}~\bibnamefont {Wagemans}}}\ (\bibinfo  {publisher} {CRC Press, Boca
  Raton, FL},\ \bibinfo {year} {1991})\BibitemShut {NoStop}%
\bibitem [{\citenamefont {Rizea}\ and\ \citenamefont {Carjan}(2013)}]{riz13}%
  \BibitemOpen
  \bibfield  {author} {\bibinfo {author} {\bibfnamefont {M.}~\bibnamefont
  {Rizea}}\ and\ \bibinfo {author} {\bibfnamefont {N.}~\bibnamefont {Carjan}},\
  }\href {\doibase http://dx.doi.org/10.1016/j.nuclphysa.2013.04.014}
  {\bibfield  {journal} {\bibinfo  {journal} {Nucl. Phys. A}\ }\textbf
  {\bibinfo {volume} {909}},\ \bibinfo {pages} {50 } (\bibinfo {year}
  {2013})}\BibitemShut {NoStop}%
\bibitem [{\citenamefont {Dubray}\ \emph {et~al.}(2008)\citenamefont {Dubray},
  \citenamefont {Goutte},\ and\ \citenamefont {Delaroche}}]{dub08}%
  \BibitemOpen
  \bibfield  {author} {\bibinfo {author} {\bibfnamefont {N.}~\bibnamefont
  {Dubray}}, \bibinfo {author} {\bibfnamefont {H.}~\bibnamefont {Goutte}}, \
  and\ \bibinfo {author} {\bibfnamefont {J.-P.}\ \bibnamefont {Delaroche}},\
  }\href {\doibase 10.1103/PhysRevC.77.014310} {\bibfield  {journal} {\bibinfo
  {journal} {Phys. Rev. C}\ }\textbf {\bibinfo {volume} {77}},\ \bibinfo
  {pages} {014310} (\bibinfo {year} {2008})}\BibitemShut {NoStop}%
\bibitem [{\citenamefont {Goutte}\ \emph {et~al.}(2005)\citenamefont {Goutte},
  \citenamefont {Berger}, \citenamefont {Casoli},\ and\ \citenamefont
  {Gogny}}]{gou05}%
  \BibitemOpen
  \bibfield  {author} {\bibinfo {author} {\bibfnamefont {H.}~\bibnamefont
  {Goutte}}, \bibinfo {author} {\bibfnamefont {J.~F.}\ \bibnamefont {Berger}},
  \bibinfo {author} {\bibfnamefont {P.}~\bibnamefont {Casoli}}, \ and\ \bibinfo
  {author} {\bibfnamefont {D.}~\bibnamefont {Gogny}},\ }\href {\doibase
  10.1103/PhysRevC.71.024316} {\bibfield  {journal} {\bibinfo  {journal} {Phys.
  Rev. C}\ }\textbf {\bibinfo {volume} {71}},\ \bibinfo {pages} {024316}
  (\bibinfo {year} {2005})}\BibitemShut {NoStop}%
\bibitem [{\citenamefont {Bonneau}(2006)}]{bon06}%
  \BibitemOpen
  \bibfield  {author} {\bibinfo {author} {\bibfnamefont {L.}~\bibnamefont
  {Bonneau}},\ }\href {\doibase 10.1103/PhysRevC.74.014301} {\bibfield
  {journal} {\bibinfo  {journal} {Phys. Rev. C}\ }\textbf {\bibinfo {volume}
  {74}},\ \bibinfo {pages} {014301} (\bibinfo {year} {2006})}\BibitemShut
  {NoStop}%
\bibitem [{\citenamefont {Staszczak}\ \emph {et~al.}(2009)\citenamefont
  {Staszczak}, \citenamefont {Baran}, \citenamefont {Dobaczewski},\ and\
  \citenamefont {Nazarewicz}}]{sta09}%
  \BibitemOpen
  \bibfield  {author} {\bibinfo {author} {\bibfnamefont {A.}~\bibnamefont
  {Staszczak}}, \bibinfo {author} {\bibfnamefont {A.}~\bibnamefont {Baran}},
  \bibinfo {author} {\bibfnamefont {J.}~\bibnamefont {Dobaczewski}}, \ and\
  \bibinfo {author} {\bibfnamefont {W.}~\bibnamefont {Nazarewicz}},\ }\href
  {\doibase 10.1103/PhysRevC.80.014309} {\bibfield  {journal} {\bibinfo
  {journal} {Phys. Rev. C}\ }\textbf {\bibinfo {volume} {80}},\ \bibinfo
  {pages} {014309} (\bibinfo {year} {2009})}\BibitemShut {NoStop}%
\bibitem [{\citenamefont {Pei}\ \emph {et~al.}(2009)\citenamefont {Pei},
  \citenamefont {Nazarewicz}, \citenamefont {Sheikh},\ and\ \citenamefont
  {Kerman}}]{pei09}%
  \BibitemOpen
  \bibfield  {author} {\bibinfo {author} {\bibfnamefont {J.~C.}\ \bibnamefont
  {Pei}}, \bibinfo {author} {\bibfnamefont {W.}~\bibnamefont {Nazarewicz}},
  \bibinfo {author} {\bibfnamefont {J.~A.}\ \bibnamefont {Sheikh}}, \ and\
  \bibinfo {author} {\bibfnamefont {A.~K.}\ \bibnamefont {Kerman}},\ }\href
  {\doibase 10.1103/PhysRevLett.102.192501} {\bibfield  {journal} {\bibinfo
  {journal} {Phys. Rev. Lett.}\ }\textbf {\bibinfo {volume} {102}},\ \bibinfo
  {pages} {192501} (\bibinfo {year} {2009})}\BibitemShut {NoStop}%
\bibitem [{\citenamefont {Younes}\ and\ \citenamefont {Gogny}(2011)}]{you11}%
  \BibitemOpen
  \bibfield  {author} {\bibinfo {author} {\bibfnamefont {W.}~\bibnamefont
  {Younes}}\ and\ \bibinfo {author} {\bibfnamefont {D.}~\bibnamefont {Gogny}},\
  }\href {\doibase 10.1103/PhysRevLett.107.132501} {\bibfield  {journal}
  {\bibinfo  {journal} {Phys. Rev. Lett.}\ }\textbf {\bibinfo {volume} {107}},\
  \bibinfo {pages} {132501} (\bibinfo {year} {2011})}\BibitemShut {NoStop}%
\bibitem [{\citenamefont {Warda}\ and\ \citenamefont {Egido}(2012)}]{war12}%
  \BibitemOpen
  \bibfield  {author} {\bibinfo {author} {\bibfnamefont {M.}~\bibnamefont
  {Warda}}\ and\ \bibinfo {author} {\bibfnamefont {J.~L.}\ \bibnamefont
  {Egido}},\ }\href {\doibase 10.1103/PhysRevC.86.014322} {\bibfield  {journal}
  {\bibinfo  {journal} {Phys. Rev. C}\ }\textbf {\bibinfo {volume} {86}},\
  \bibinfo {pages} {014322} (\bibinfo {year} {2012})}\BibitemShut {NoStop}%
\bibitem [{\citenamefont {Abusara}\ \emph {et~al.}(2012)\citenamefont
  {Abusara}, \citenamefont {Afanasjev},\ and\ \citenamefont {Ring}}]{abu12}%
  \BibitemOpen
  \bibfield  {author} {\bibinfo {author} {\bibfnamefont {H.}~\bibnamefont
  {Abusara}}, \bibinfo {author} {\bibfnamefont {A.~V.}\ \bibnamefont
  {Afanasjev}}, \ and\ \bibinfo {author} {\bibfnamefont {P.}~\bibnamefont
  {Ring}},\ }\href {\doibase 10.1103/PhysRevC.85.024314} {\bibfield  {journal}
  {\bibinfo  {journal} {Phys. Rev. C}\ }\textbf {\bibinfo {volume} {85}},\
  \bibinfo {pages} {024314} (\bibinfo {year} {2012})}\BibitemShut {NoStop}%
\bibitem [{\citenamefont {Mirea}(2012)}]{mir12}%
  \BibitemOpen
  \bibfield  {author} {\bibinfo {author} {\bibfnamefont {M.}~\bibnamefont
  {Mirea}},\ }\href {\doibase http://dx.doi.org/10.1016/j.physletb.2012.09.023}
  {\bibfield  {journal} {\bibinfo  {journal} {Phys. Lett. B}\ }\textbf
  {\bibinfo {volume} {717}},\ \bibinfo {pages} {252 } (\bibinfo {year}
  {2012})}\BibitemShut {NoStop}%
\bibitem [{\citenamefont {Staszczak}\ \emph {et~al.}(2013)\citenamefont
  {Staszczak}, \citenamefont {Baran},\ and\ \citenamefont
  {Nazarewicz}}]{sta13}%
  \BibitemOpen
  \bibfield  {author} {\bibinfo {author} {\bibfnamefont {A.}~\bibnamefont
  {Staszczak}}, \bibinfo {author} {\bibfnamefont {A.}~\bibnamefont {Baran}}, \
  and\ \bibinfo {author} {\bibfnamefont {W.}~\bibnamefont {Nazarewicz}},\
  }\href {\doibase 10.1103/PhysRevC.87.024320} {\bibfield  {journal} {\bibinfo
  {journal} {Phys. Rev. C}\ }\textbf {\bibinfo {volume} {87}},\ \bibinfo
  {pages} {024320} (\bibinfo {year} {2013})}\BibitemShut {NoStop}%
\bibitem [{\citenamefont {McDonnell}\ \emph {et~al.}(2013)\citenamefont
  {McDonnell}, \citenamefont {Nazarewicz},\ and\ \citenamefont
  {Sheikh}}]{mcd13}%
  \BibitemOpen
  \bibfield  {author} {\bibinfo {author} {\bibfnamefont {J.~D.}\ \bibnamefont
  {McDonnell}}, \bibinfo {author} {\bibfnamefont {W.}~\bibnamefont
  {Nazarewicz}}, \ and\ \bibinfo {author} {\bibfnamefont {J.~A.}\ \bibnamefont
  {Sheikh}},\ }\href {\doibase 10.1103/PhysRevC.87.054327} {\bibfield
  {journal} {\bibinfo  {journal} {Phys. Rev. C}\ }\textbf {\bibinfo {volume}
  {87}},\ \bibinfo {pages} {054327} (\bibinfo {year} {2013})}\BibitemShut
  {NoStop}%
\bibitem [{\citenamefont {Sadhukhan}\ \emph {et~al.}(2013)\citenamefont
  {Sadhukhan}, \citenamefont {Mazurek}, \citenamefont {Baran}, \citenamefont
  {Dobaczewski}, \citenamefont {Nazarewicz},\ and\ \citenamefont
  {Sheikh}}]{sad13}%
  \BibitemOpen
  \bibfield  {author} {\bibinfo {author} {\bibfnamefont {J.}~\bibnamefont
  {Sadhukhan}}, \bibinfo {author} {\bibfnamefont {K.}~\bibnamefont {Mazurek}},
  \bibinfo {author} {\bibfnamefont {A.}~\bibnamefont {Baran}}, \bibinfo
  {author} {\bibfnamefont {J.}~\bibnamefont {Dobaczewski}}, \bibinfo {author}
  {\bibfnamefont {W.}~\bibnamefont {Nazarewicz}}, \ and\ \bibinfo {author}
  {\bibfnamefont {J.~A.}\ \bibnamefont {Sheikh}},\ }\href {\doibase
  10.1103/PhysRevC.88.064314} {\bibfield  {journal} {\bibinfo  {journal} {Phys.
  Rev. C}\ }\textbf {\bibinfo {volume} {88}},\ \bibinfo {pages} {064314}
  (\bibinfo {year} {2013})}\BibitemShut {NoStop}%
\bibitem [{\citenamefont {Schunck}\ \emph {et~al.}({\natexlab{a}})\citenamefont
  {Schunck}, \citenamefont {Duke}, \citenamefont {Carr},\ and\ \citenamefont
  {Knoll}}]{sch14a}%
  \BibitemOpen
  \bibfield  {author} {\bibinfo {author} {\bibfnamefont {N.}~\bibnamefont
  {Schunck}}, \bibinfo {author} {\bibfnamefont {D.}~\bibnamefont {Duke}},
  \bibinfo {author} {\bibfnamefont {H.}~\bibnamefont {Carr}}, \ and\ \bibinfo
  {author} {\bibfnamefont {A.}~\bibnamefont {Knoll}},\ }\href@noop {} {}
  ({\natexlab{a}}),\ \bibinfo {note} {arXiv:1311.2616}\BibitemShut {NoStop}%
\bibitem [{\citenamefont {Schunck}\ \emph {et~al.}({\natexlab{b}})\citenamefont
  {Schunck}, \citenamefont {Duke},\ and\ \citenamefont {Carr}}]{sch14b}%
  \BibitemOpen
  \bibfield  {author} {\bibinfo {author} {\bibfnamefont {N.}~\bibnamefont
  {Schunck}}, \bibinfo {author} {\bibfnamefont {D.}~\bibnamefont {Duke}}, \
  and\ \bibinfo {author} {\bibfnamefont {H.}~\bibnamefont {Carr}},\ }\href@noop
  {} {} ({\natexlab{b}}),\ \bibinfo {note} {arXiv:1311.2620}\BibitemShut
  {NoStop}%
\bibitem [{\citenamefont {Baranger}\ and\ \citenamefont
  {V\'en\'eroni}(1978)}]{bar78}%
  \BibitemOpen
  \bibfield  {author} {\bibinfo {author} {\bibfnamefont {M.}~\bibnamefont
  {Baranger}}\ and\ \bibinfo {author} {\bibfnamefont {M.}~\bibnamefont
  {V\'en\'eroni}},\ }\href@noop {} {\bibfield  {journal} {\bibinfo  {journal}
  {Ann. Phys. (New York)}\ }\textbf {\bibinfo {volume} {114}},\ \bibinfo
  {pages} {123} (\bibinfo {year} {1978})}\BibitemShut {NoStop}%
\bibitem [{\citenamefont {M\"oller}\ \emph {et~al.}(2001)\citenamefont
  {M\"oller}, \citenamefont {Madland}, \citenamefont {Sierk},\ and\
  \citenamefont {Iwamoto}}]{mol01}%
  \BibitemOpen
  \bibfield  {author} {\bibinfo {author} {\bibfnamefont {P.}~\bibnamefont
  {M\"oller}}, \bibinfo {author} {\bibfnamefont {D.~G.}\ \bibnamefont
  {Madland}}, \bibinfo {author} {\bibfnamefont {A.~J.}\ \bibnamefont {Sierk}},
  \ and\ \bibinfo {author} {\bibfnamefont {A.}~\bibnamefont {Iwamoto}},\ }\href
  {\doibase 10.1038/35057204} {\bibfield  {journal} {\bibinfo  {journal}
  {Nature}\ }\textbf {\bibinfo {volume} {409}},\ \bibinfo {pages} {785}
  (\bibinfo {year} {2001})}\BibitemShut {NoStop}%
\bibitem [{\citenamefont {Randrup}\ and\ \citenamefont
  {M\"oller}(2011)}]{ran11}%
  \BibitemOpen
  \bibfield  {author} {\bibinfo {author} {\bibfnamefont {J.}~\bibnamefont
  {Randrup}}\ and\ \bibinfo {author} {\bibfnamefont {P.}~\bibnamefont
  {M\"oller}},\ }\href {\doibase 10.1103/PhysRevLett.106.132503} {\bibfield
  {journal} {\bibinfo  {journal} {Phys. Rev. Lett.}\ }\textbf {\bibinfo
  {volume} {106}},\ \bibinfo {pages} {132503} (\bibinfo {year}
  {2011})}\BibitemShut {NoStop}%
\bibitem [{\citenamefont {Wilkins}\ \emph {et~al.}(1976)\citenamefont
  {Wilkins}, \citenamefont {Steinberg},\ and\ \citenamefont {Chasman}}]{wil76}%
  \BibitemOpen
  \bibfield  {author} {\bibinfo {author} {\bibfnamefont {B.~D.}\ \bibnamefont
  {Wilkins}}, \bibinfo {author} {\bibfnamefont {E.~P.}\ \bibnamefont
  {Steinberg}}, \ and\ \bibinfo {author} {\bibfnamefont {R.~R.}\ \bibnamefont
  {Chasman}},\ }\href {\doibase 10.1103/PhysRevC.14.1832} {\bibfield  {journal}
  {\bibinfo  {journal} {Phys. Rev. C}\ }\textbf {\bibinfo {volume} {14}},\
  \bibinfo {pages} {1832} (\bibinfo {year} {1976})}\BibitemShut {NoStop}%
\bibitem [{\citenamefont {Bernard}\ \emph {et~al.}(2011)\citenamefont
  {Bernard}, \citenamefont {Goutte}, \citenamefont {Gogny},\ and\ \citenamefont
  {Younes}}]{ber11}%
  \BibitemOpen
  \bibfield  {author} {\bibinfo {author} {\bibfnamefont {R.}~\bibnamefont
  {Bernard}}, \bibinfo {author} {\bibfnamefont {H.}~\bibnamefont {Goutte}},
  \bibinfo {author} {\bibfnamefont {D.}~\bibnamefont {Gogny}}, \ and\ \bibinfo
  {author} {\bibfnamefont {W.}~\bibnamefont {Younes}},\ }\href {\doibase
  10.1103/PhysRevC.84.044308} {\bibfield  {journal} {\bibinfo  {journal} {Phys.
  Rev. C}\ }\textbf {\bibinfo {volume} {84}},\ \bibinfo {pages} {044308}
  (\bibinfo {year} {2011})}\BibitemShut {NoStop}%
\bibitem [{\citenamefont {Umar}\ \emph {et~al.}(2010)\citenamefont {Umar},
  \citenamefont {Oberacker}, \citenamefont {Maruhn},\ and\ \citenamefont
  {Reinhard}}]{uma10c}%
  \BibitemOpen
  \bibfield  {author} {\bibinfo {author} {\bibfnamefont {A.~S.}\ \bibnamefont
  {Umar}}, \bibinfo {author} {\bibfnamefont {V.~E.}\ \bibnamefont {Oberacker}},
  \bibinfo {author} {\bibfnamefont {J.~A.}\ \bibnamefont {Maruhn}}, \ and\
  \bibinfo {author} {\bibfnamefont {P.-G.}\ \bibnamefont {Reinhard}},\ }\href
  {http://stacks.iop.org/0954-3899/37/i=6/a=064037} {\bibfield  {journal}
  {\bibinfo  {journal} {J. Phys. G}\ }\textbf {\bibinfo {volume} {37}},\
  \bibinfo {pages} {064037} (\bibinfo {year} {2010})}\BibitemShut {NoStop}%
\bibitem [{\citenamefont {Dirac}(1930)}]{dir30}%
  \BibitemOpen
  \bibfield  {author} {\bibinfo {author} {\bibfnamefont {P.~A.~M.}\
  \bibnamefont {Dirac}},\ }\href@noop {} {\bibfield  {journal} {\bibinfo
  {journal} {Proc. Camb. Phil. Soc.}\ }\textbf {\bibinfo {volume} {26}},\
  \bibinfo {pages} {376} (\bibinfo {year} {1930})}\BibitemShut {NoStop}%
\bibitem [{\citenamefont {Negele}\ \emph {et~al.}(1978)\citenamefont {Negele},
  \citenamefont {Koonin}, \citenamefont {M\"oller}, \citenamefont {Nix},\ and\
  \citenamefont {Sierk}}]{neg78}%
  \BibitemOpen
  \bibfield  {author} {\bibinfo {author} {\bibfnamefont {J.~W.}\ \bibnamefont
  {Negele}}, \bibinfo {author} {\bibfnamefont {S.~E.}\ \bibnamefont {Koonin}},
  \bibinfo {author} {\bibfnamefont {P.}~\bibnamefont {M\"oller}}, \bibinfo
  {author} {\bibfnamefont {J.~R.}\ \bibnamefont {Nix}}, \ and\ \bibinfo
  {author} {\bibfnamefont {A.~J.}\ \bibnamefont {Sierk}},\ }\href {\doibase
  10.1103/PhysRevC.17.1098} {\bibfield  {journal} {\bibinfo  {journal} {Phys.
  Rev. C}\ }\textbf {\bibinfo {volume} {17}},\ \bibinfo {pages} {1098}
  (\bibinfo {year} {1978})}\BibitemShut {NoStop}%
\bibitem [{\citenamefont {M\"oller}\ and\ \citenamefont
  {Iwamoto}(2000)}]{mol00}%
  \BibitemOpen
  \bibfield  {author} {\bibinfo {author} {\bibfnamefont {P.}~\bibnamefont
  {M\"oller}}\ and\ \bibinfo {author} {\bibfnamefont {A.}~\bibnamefont
  {Iwamoto}},\ }\href {\doibase 10.1103/PhysRevC.61.047602} {\bibfield
  {journal} {\bibinfo  {journal} {Phys. Rev. C}\ }\textbf {\bibinfo {volume}
  {61}},\ \bibinfo {pages} {047602} (\bibinfo {year} {2000})}\BibitemShut
  {NoStop}%
\bibitem [{\citenamefont {Gherghescu}\ \emph {et~al.}(1996)\citenamefont
  {Gherghescu}, \citenamefont {Poenaru},\ and\ \citenamefont
  {Greiner}}]{ghe96}%
  \BibitemOpen
  \bibfield  {author} {\bibinfo {author} {\bibfnamefont {R.}~\bibnamefont
  {Gherghescu}}, \bibinfo {author} {\bibfnamefont {D.}~\bibnamefont {Poenaru}},
  \ and\ \bibinfo {author} {\bibfnamefont {W.}~\bibnamefont {Greiner}},\ }\href
  {\doibase 10.1007/BF02769540} {\bibfield  {journal} {\bibinfo  {journal} {Z.
  Phys. A}\ }\textbf {\bibinfo {volume} {354}},\ \bibinfo {pages} {367}
  (\bibinfo {year} {1996})}\BibitemShut {NoStop}%
\bibitem [{\citenamefont {Asano}\ \emph {et~al.}(2004)\citenamefont {Asano},
  \citenamefont {Wada}, \citenamefont {Ohta}, \citenamefont {Ichikawa},
  \citenamefont {Yamaji},\ and\ \citenamefont {Nakahara}}]{asa04}%
  \BibitemOpen
  \bibfield  {author} {\bibinfo {author} {\bibfnamefont {T.}~\bibnamefont
  {Asano}}, \bibinfo {author} {\bibfnamefont {T.}~\bibnamefont {Wada}},
  \bibinfo {author} {\bibfnamefont {M.}~\bibnamefont {Ohta}}, \bibinfo {author}
  {\bibfnamefont {T.}~\bibnamefont {Ichikawa}}, \bibinfo {author}
  {\bibfnamefont {S.}~\bibnamefont {Yamaji}}, \ and\ \bibinfo {author}
  {\bibfnamefont {H.}~\bibnamefont {Nakahara}},\ }\href@noop {} {\bibfield
  {journal} {\bibinfo  {journal} {J. Nucl. Radiochem. Sci.}\ }\textbf {\bibinfo
  {volume} {5}},\ \bibinfo {pages} {1} (\bibinfo {year} {2004})}\BibitemShut
  {NoStop}%
\bibitem [{\citenamefont {Dvorak}\ \emph {et~al.}(2011)\citenamefont {Dvorak},
  \citenamefont {Block}, \citenamefont {D{\"u}llmann}, \citenamefont {Heinz},
  \citenamefont {Herzberg},\ and\ \citenamefont {Sch{\"a}del}}]{dvo11}%
  \BibitemOpen
  \bibfield  {author} {\bibinfo {author} {\bibfnamefont {J.}~\bibnamefont
  {Dvorak}}, \bibinfo {author} {\bibfnamefont {M.}~\bibnamefont {Block}},
  \bibinfo {author} {\bibfnamefont {C.}~\bibnamefont {D{\"u}llmann}}, \bibinfo
  {author} {\bibfnamefont {S.}~\bibnamefont {Heinz}}, \bibinfo {author}
  {\bibfnamefont {R.-D.}\ \bibnamefont {Herzberg}}, \ and\ \bibinfo {author}
  {\bibfnamefont {M.}~\bibnamefont {Sch{\"a}del}},\ }\href {\doibase
  10.1016/j.nima.2010.08.124} {\bibfield  {journal} {\bibinfo  {journal} {Nucl.
  Instr. Meth. A}\ }\textbf {\bibinfo {volume} {652}},\ \bibinfo {pages} {687}
  (\bibinfo {year} {2011})}\BibitemShut {NoStop}%
\bibitem [{\citenamefont {Zagrebaev}\ \emph {et~al.}(2006)\citenamefont
  {Zagrebaev}, \citenamefont {Oganessian}, \citenamefont {Itkis},\ and\
  \citenamefont {Greiner}}]{zag06}%
  \BibitemOpen
  \bibfield  {author} {\bibinfo {author} {\bibfnamefont {V.~I.}\ \bibnamefont
  {Zagrebaev}}, \bibinfo {author} {\bibfnamefont {Y.~T.}\ \bibnamefont
  {Oganessian}}, \bibinfo {author} {\bibfnamefont {M.~G.}\ \bibnamefont
  {Itkis}}, \ and\ \bibinfo {author} {\bibfnamefont {W.}~\bibnamefont
  {Greiner}},\ }\href {\doibase 10.1103/PhysRevC.73.031602} {\bibfield
  {journal} {\bibinfo  {journal} {Phys. Rev. C}\ }\textbf {\bibinfo {volume}
  {73}},\ \bibinfo {pages} {031602} (\bibinfo {year} {2006})}\BibitemShut
  {NoStop}%
\bibitem [{\citenamefont {Golabek}\ and\ \citenamefont
  {Simenel}(2009)}]{gol09}%
  \BibitemOpen
  \bibfield  {author} {\bibinfo {author} {\bibfnamefont {C.}~\bibnamefont
  {Golabek}}\ and\ \bibinfo {author} {\bibfnamefont {C.}~\bibnamefont
  {Simenel}},\ }\href {\doibase 10.1103/PhysRevLett.103.042701} {\bibfield
  {journal} {\bibinfo  {journal} {Phys. Rev. Lett.}\ }\textbf {\bibinfo
  {volume} {103}},\ \bibinfo {pages} {042701} (\bibinfo {year}
  {2009})}\BibitemShut {NoStop}%
\bibitem [{\citenamefont {Kedziora}\ and\ \citenamefont
  {Simenel}(2010)}]{ked10}%
  \BibitemOpen
  \bibfield  {author} {\bibinfo {author} {\bibfnamefont {D.~J.}\ \bibnamefont
  {Kedziora}}\ and\ \bibinfo {author} {\bibfnamefont {C.}~\bibnamefont
  {Simenel}},\ }\href {\doibase 10.1103/PhysRevC.81.044613} {\bibfield
  {journal} {\bibinfo  {journal} {Phys. Rev. C}\ }\textbf {\bibinfo {volume}
  {81}},\ \bibinfo {pages} {044613} (\bibinfo {year} {2010})}\BibitemShut
  {NoStop}%
\bibitem [{\citenamefont {Bonche}\ \emph {et~al.}(2005)\citenamefont {Bonche},
  \citenamefont {Flocard},\ and\ \citenamefont {Heenen}}]{bon05}%
  \BibitemOpen
  \bibfield  {author} {\bibinfo {author} {\bibfnamefont {P.}~\bibnamefont
  {Bonche}}, \bibinfo {author} {\bibfnamefont {H.}~\bibnamefont {Flocard}}, \
  and\ \bibinfo {author} {\bibfnamefont {P.~H.}\ \bibnamefont {Heenen}},\
  }\href {\doibase 10.1016/j.cpc.2005.05.001} {\bibfield  {journal} {\bibinfo
  {journal} {Comp. Phys. Com.}\ }\textbf {\bibinfo {volume} {171}},\ \bibinfo
  {pages} {49} (\bibinfo {year} {2005})}\BibitemShut {NoStop}%
\bibitem [{\citenamefont {Kim}\ \emph {et~al.}(1997)\citenamefont {Kim},
  \citenamefont {Otsuka},\ and\ \citenamefont {Bonche}}]{kim97}%
  \BibitemOpen
  \bibfield  {author} {\bibinfo {author} {\bibfnamefont {K.-H.}\ \bibnamefont
  {Kim}}, \bibinfo {author} {\bibfnamefont {T.}~\bibnamefont {Otsuka}}, \ and\
  \bibinfo {author} {\bibfnamefont {P.}~\bibnamefont {Bonche}},\ }\href@noop {}
  {\bibfield  {journal} {\bibinfo  {journal} {J. Phys. G}\ }\textbf {\bibinfo
  {volume} {23}},\ \bibinfo {pages} {1267} (\bibinfo {year}
  {1997})}\BibitemShut {NoStop}%
\bibitem [{\citenamefont {Skyrme}(1956)}]{sky56}%
  \BibitemOpen
  \bibfield  {author} {\bibinfo {author} {\bibfnamefont {T.}~\bibnamefont
  {Skyrme}},\ }\href@noop {} {\bibfield  {journal} {\bibinfo  {journal} {Phil.
  Mag.}\ }\textbf {\bibinfo {volume} {1}},\ \bibinfo {pages} {1043} (\bibinfo
  {year} {1956})}\BibitemShut {NoStop}%
\bibitem [{\citenamefont {Bender}\ \emph {et~al.}(2003)\citenamefont {Bender},
  \citenamefont {Heenen},\ and\ \citenamefont {Reinhard}}]{bender03}%
  \BibitemOpen
  \bibfield  {author} {\bibinfo {author} {\bibfnamefont {M.}~\bibnamefont
  {Bender}}, \bibinfo {author} {\bibfnamefont {P.-H.}\ \bibnamefont {Heenen}},
  \ and\ \bibinfo {author} {\bibfnamefont {P.-G.}\ \bibnamefont {Reinhard}},\
  }\href {\doibase 10.1103/RevModPhys.75.121} {\bibfield  {journal} {\bibinfo
  {journal} {Rev. Mod. Phys.}\ }\textbf {\bibinfo {volume} {75}},\ \bibinfo
  {pages} {121} (\bibinfo {year} {2003})}\BibitemShut {NoStop}%
\bibitem [{\citenamefont {M\"oller}\ \emph {et~al.}(2009)\citenamefont
  {M\"oller}, \citenamefont {Sierk}, \citenamefont {Ichikawa}, \citenamefont
  {Iwamoto}, \citenamefont {Bengtsson}, \citenamefont {Uhrenholt},\ and\
  \citenamefont {\AA{}berg}}]{mol09}%
  \BibitemOpen
  \bibfield  {author} {\bibinfo {author} {\bibfnamefont {P.}~\bibnamefont
  {M\"oller}}, \bibinfo {author} {\bibfnamefont {A.~J.}\ \bibnamefont {Sierk}},
  \bibinfo {author} {\bibfnamefont {T.}~\bibnamefont {Ichikawa}}, \bibinfo
  {author} {\bibfnamefont {A.}~\bibnamefont {Iwamoto}}, \bibinfo {author}
  {\bibfnamefont {R.}~\bibnamefont {Bengtsson}}, \bibinfo {author}
  {\bibfnamefont {H.}~\bibnamefont {Uhrenholt}}, \ and\ \bibinfo {author}
  {\bibfnamefont {S.}~\bibnamefont {\AA{}berg}},\ }\href {\doibase
  10.1103/PhysRevC.79.064304} {\bibfield  {journal} {\bibinfo  {journal} {Phys.
  Rev. C}\ }\textbf {\bibinfo {volume} {79}},\ \bibinfo {pages} {064304}
  (\bibinfo {year} {2009})}\BibitemShut {NoStop}%
\bibitem [{\citenamefont {Kowal}\ \emph {et~al.}(2010)\citenamefont {Kowal},
  \citenamefont {Jachimowicz},\ and\ \citenamefont {Sobiczewski}}]{kow10}%
  \BibitemOpen
  \bibfield  {author} {\bibinfo {author} {\bibfnamefont {M.}~\bibnamefont
  {Kowal}}, \bibinfo {author} {\bibfnamefont {P.}~\bibnamefont {Jachimowicz}},
  \ and\ \bibinfo {author} {\bibfnamefont {A.}~\bibnamefont {Sobiczewski}},\
  }\href {\doibase 10.1103/PhysRevC.82.014303} {\bibfield  {journal} {\bibinfo
  {journal} {Phys. Rev. C}\ }\textbf {\bibinfo {volume} {82}},\ \bibinfo
  {pages} {014303} (\bibinfo {year} {2010})}\BibitemShut {NoStop}%
\bibitem [{\citenamefont {Hulet}\ \emph {et~al.}(1986)\citenamefont {Hulet},
  \citenamefont {Wild}, \citenamefont {Dougan}, \citenamefont {Lougheed},
  \citenamefont {Landrum}, \citenamefont {Dougan}, \citenamefont {Schadel},
  \citenamefont {Hahn}, \citenamefont {Baisden}, \citenamefont {Henderson},
  \citenamefont {Dupzyk}, \citenamefont {S\"ummerer},\ and\ \citenamefont
  {Bethune}}]{hul86}%
  \BibitemOpen
  \bibfield  {author} {\bibinfo {author} {\bibfnamefont {E.~K.}\ \bibnamefont
  {Hulet}}, \bibinfo {author} {\bibfnamefont {J.~F.}\ \bibnamefont {Wild}},
  \bibinfo {author} {\bibfnamefont {R.~J.}\ \bibnamefont {Dougan}}, \bibinfo
  {author} {\bibfnamefont {R.~W.}\ \bibnamefont {Lougheed}}, \bibinfo {author}
  {\bibfnamefont {J.~H.}\ \bibnamefont {Landrum}}, \bibinfo {author}
  {\bibfnamefont {A.~D.}\ \bibnamefont {Dougan}}, \bibinfo {author}
  {\bibfnamefont {M.}~\bibnamefont {Schadel}}, \bibinfo {author} {\bibfnamefont
  {R.~L.}\ \bibnamefont {Hahn}}, \bibinfo {author} {\bibfnamefont {P.~A.}\
  \bibnamefont {Baisden}}, \bibinfo {author} {\bibfnamefont {C.~M.}\
  \bibnamefont {Henderson}}, \bibinfo {author} {\bibfnamefont {R.~J.}\
  \bibnamefont {Dupzyk}}, \bibinfo {author} {\bibfnamefont {K.}~\bibnamefont
  {S\"ummerer}}, \ and\ \bibinfo {author} {\bibfnamefont {G.~R.}\ \bibnamefont
  {Bethune}},\ }\href {\doibase 10.1103/PhysRevLett.56.313} {\bibfield
  {journal} {\bibinfo  {journal} {Phys. Rev. Lett.}\ }\textbf {\bibinfo
  {volume} {56}},\ \bibinfo {pages} {313} (\bibinfo {year} {1986})}\BibitemShut
  {NoStop}%
\bibitem [{\citenamefont {Ichikawa}\ \emph {et~al.}(2009)\citenamefont
  {Ichikawa}, \citenamefont {Iwamoto},\ and\ \citenamefont {M\"oller}}]{ich09}%
  \BibitemOpen
  \bibfield  {author} {\bibinfo {author} {\bibfnamefont {T.}~\bibnamefont
  {Ichikawa}}, \bibinfo {author} {\bibfnamefont {A.}~\bibnamefont {Iwamoto}}, \
  and\ \bibinfo {author} {\bibfnamefont {P.}~\bibnamefont {M\"oller}},\ }\href
  {\doibase 10.1103/PhysRevC.79.014305} {\bibfield  {journal} {\bibinfo
  {journal} {Phys. Rev. C}\ }\textbf {\bibinfo {volume} {79}},\ \bibinfo
  {pages} {014305} (\bibinfo {year} {2009})}\BibitemShut {NoStop}%
\bibitem [{\citenamefont {Ring}\ and\ \citenamefont {Schuck}(1980)}]{rin80}%
  \BibitemOpen
  \bibfield  {author} {\bibinfo {author} {\bibfnamefont {P.}~\bibnamefont
  {Ring}}\ and\ \bibinfo {author} {\bibfnamefont {P.}~\bibnamefont {Schuck}},\
  }\href@noop {} {\emph {\bibinfo {title} {The Nuclear Many-Body Problem}}}\
  (\bibinfo  {publisher} {Springer Verlag},\ \bibinfo {year}
  {1980})\BibitemShut {NoStop}%
\bibitem [{\citenamefont {Negele}(1982)}]{neg82}%
  \BibitemOpen
  \bibfield  {author} {\bibinfo {author} {\bibfnamefont {J.~W.}\ \bibnamefont
  {Negele}},\ }\href {\doibase 10.1103/RevModPhys.54.913} {\bibfield  {journal}
  {\bibinfo  {journal} {Rev. Mod. Phys.}\ }\textbf {\bibinfo {volume} {54}},\
  \bibinfo {pages} {913} (\bibinfo {year} {1982})}\BibitemShut {NoStop}%
\bibitem [{\citenamefont {Simenel}(2012)}]{sim12b}%
  \BibitemOpen
  \bibfield  {author} {\bibinfo {author} {\bibfnamefont {C.}~\bibnamefont
  {Simenel}},\ }\href {\doibase 10.1140/epja/i2012-12152-0} {\bibfield
  {journal} {\bibinfo  {journal} {Eur. Phys. J. A}\ }\textbf {\bibinfo {volume}
  {48}},\ \bibinfo {pages} {152} (\bibinfo {year} {2012})}\BibitemShut
  {NoStop}%
\bibitem [{\citenamefont {Simenel}(2014)}]{sim13a}%
  \BibitemOpen
  \bibfield  {author} {\bibinfo {author} {\bibfnamefont {C.}~\bibnamefont
  {Simenel}},\ }in\ \href {\doibase 10.1007/978-3-319-01077-9_4} {\emph
  {\bibinfo {booktitle} {Clusters in Nuclei, Vol. 3}}},\ \bibinfo {series}
  {Lecture Notes in Physics}, Vol.\ \bibinfo {volume} {875},\ \bibinfo {editor}
  {edited by\ \bibinfo {editor} {\bibfnamefont {C.}~\bibnamefont {Beck}}}\
  (\bibinfo  {publisher} {Springer International Publishing},\ \bibinfo {year}
  {2014})\ pp.\ \bibinfo {pages} {95--145}\BibitemShut {NoStop}%
\bibitem [{\citenamefont {B{\l}ocki}\ and\ \citenamefont
  {Flocard}(1976)}]{blo76}%
  \BibitemOpen
  \bibfield  {author} {\bibinfo {author} {\bibfnamefont {J.}~\bibnamefont
  {B{\l}ocki}}\ and\ \bibinfo {author} {\bibfnamefont {H.}~\bibnamefont
  {Flocard}},\ }\href {\doibase http://dx.doi.org/10.1016/0375-9474(76)90299-2}
  {\bibfield  {journal} {\bibinfo  {journal} {Nucl. Phys. A}\ }\textbf
  {\bibinfo {volume} {273}},\ \bibinfo {pages} {45 } (\bibinfo {year}
  {1976})}\BibitemShut {NoStop}%
\bibitem [{\citenamefont {Avez}\ \emph {et~al.}(2008)\citenamefont {Avez},
  \citenamefont {Simenel},\ and\ \citenamefont {Chomaz}}]{ave08}%
  \BibitemOpen
  \bibfield  {author} {\bibinfo {author} {\bibfnamefont {B.}~\bibnamefont
  {Avez}}, \bibinfo {author} {\bibfnamefont {C.}~\bibnamefont {Simenel}}, \
  and\ \bibinfo {author} {\bibfnamefont {P.}~\bibnamefont {Chomaz}},\ }\href
  {\doibase 10.1103/PhysRevC.78.044318} {\bibfield  {journal} {\bibinfo
  {journal} {Phys. Rev. C}\ }\textbf {\bibinfo {volume} {78}},\ \bibinfo {eid}
  {044318} (\bibinfo {year} {2008})}\BibitemShut {NoStop}%
\bibitem [{\citenamefont {Ebata}\ \emph {et~al.}(2010)\citenamefont {Ebata},
  \citenamefont {Nakatsukasa}, \citenamefont {Inakura}, \citenamefont
  {Yoshida}, \citenamefont {Hashimoto},\ and\ \citenamefont {Yabana}}]{eba10}%
  \BibitemOpen
  \bibfield  {author} {\bibinfo {author} {\bibfnamefont {S.}~\bibnamefont
  {Ebata}}, \bibinfo {author} {\bibfnamefont {T.}~\bibnamefont {Nakatsukasa}},
  \bibinfo {author} {\bibfnamefont {T.}~\bibnamefont {Inakura}}, \bibinfo
  {author} {\bibfnamefont {K.}~\bibnamefont {Yoshida}}, \bibinfo {author}
  {\bibfnamefont {Y.}~\bibnamefont {Hashimoto}}, \ and\ \bibinfo {author}
  {\bibfnamefont {K.}~\bibnamefont {Yabana}},\ }\href {\doibase
  10.1103/PhysRevC.82.034306} {\bibfield  {journal} {\bibinfo  {journal} {Phys.
  Rev. C}\ }\textbf {\bibinfo {volume} {82}},\ \bibinfo {pages} {034306}
  (\bibinfo {year} {2010})}\BibitemShut {NoStop}%
\bibitem [{\citenamefont {Stetcu}\ \emph {et~al.}(2011)\citenamefont {Stetcu},
  \citenamefont {Bulgac}, \citenamefont {Magierski},\ and\ \citenamefont
  {Roche}}]{ste11}%
  \BibitemOpen
  \bibfield  {author} {\bibinfo {author} {\bibfnamefont {I.}~\bibnamefont
  {Stetcu}}, \bibinfo {author} {\bibfnamefont {A.}~\bibnamefont {Bulgac}},
  \bibinfo {author} {\bibfnamefont {P.}~\bibnamefont {Magierski}}, \ and\
  \bibinfo {author} {\bibfnamefont {K.~J.}\ \bibnamefont {Roche}},\ }\href
  {\doibase 10.1103/PhysRevC.84.051309} {\bibfield  {journal} {\bibinfo
  {journal} {Phys. Rev. C}\ }\textbf {\bibinfo {volume} {84}},\ \bibinfo
  {pages} {051309} (\bibinfo {year} {2011})}\BibitemShut {NoStop}%
\bibitem [{\citenamefont {Scamps}\ and\ \citenamefont {Lacroix}(2013)}]{sca13}%
  \BibitemOpen
  \bibfield  {author} {\bibinfo {author} {\bibfnamefont {G.}~\bibnamefont
  {Scamps}}\ and\ \bibinfo {author} {\bibfnamefont {D.}~\bibnamefont
  {Lacroix}},\ }\href {\doibase 10.1103/PhysRevC.87.014605} {\bibfield
  {journal} {\bibinfo  {journal} {Phys. Rev. C}\ }\textbf {\bibinfo {volume}
  {87}},\ \bibinfo {pages} {014605} (\bibinfo {year} {2013})}\BibitemShut
  {NoStop}%
\bibitem [{\citenamefont {Viola}\ \emph {et~al.}(1985)\citenamefont {Viola},
  \citenamefont {Kwiatkowski},\ and\ \citenamefont {Walker}}]{vio85}%
  \BibitemOpen
  \bibfield  {author} {\bibinfo {author} {\bibfnamefont {V.~E.}\ \bibnamefont
  {Viola}}, \bibinfo {author} {\bibfnamefont {K.}~\bibnamefont {Kwiatkowski}},
  \ and\ \bibinfo {author} {\bibfnamefont {M.}~\bibnamefont {Walker}},\ }\href
  {\doibase 10.1103/PhysRevC.31.1550} {\bibfield  {journal} {\bibinfo
  {journal} {Phys. Rev. C}\ }\textbf {\bibinfo {volume} {31}},\ \bibinfo
  {pages} {1550} (\bibinfo {year} {1985})}\BibitemShut {NoStop}%
\bibitem [{\citenamefont {Bonche}\ \emph {et~al.}(1976)\citenamefont {Bonche},
  \citenamefont {Koonin},\ and\ \citenamefont {Negele}}]{bon76}%
  \BibitemOpen
  \bibfield  {author} {\bibinfo {author} {\bibfnamefont {P.}~\bibnamefont
  {Bonche}}, \bibinfo {author} {\bibfnamefont {S.}~\bibnamefont {Koonin}}, \
  and\ \bibinfo {author} {\bibfnamefont {J.~W.}\ \bibnamefont {Negele}},\
  }\href@noop {} {\bibfield  {journal} {\bibinfo  {journal} {Phys. Rev. C}\
  }\textbf {\bibinfo {volume} {13}},\ \bibinfo {pages} {1226} (\bibinfo {year}
  {1976})}\BibitemShut {NoStop}%
\bibitem [{\citenamefont {Koonin}\ \emph {et~al.}(1977)\citenamefont {Koonin},
  \citenamefont {Davies}, \citenamefont {Maruhn-Rezwani}, \citenamefont
  {Feldmeier}, \citenamefont {Krieger},\ and\ \citenamefont {Negele}}]{koo77}%
  \BibitemOpen
  \bibfield  {author} {\bibinfo {author} {\bibfnamefont {S.~E.}\ \bibnamefont
  {Koonin}}, \bibinfo {author} {\bibfnamefont {K.~T.~R.}\ \bibnamefont
  {Davies}}, \bibinfo {author} {\bibfnamefont {V.}~\bibnamefont
  {Maruhn-Rezwani}}, \bibinfo {author} {\bibfnamefont {H.}~\bibnamefont
  {Feldmeier}}, \bibinfo {author} {\bibfnamefont {S.~J.}\ \bibnamefont
  {Krieger}}, \ and\ \bibinfo {author} {\bibfnamefont {J.~W.}\ \bibnamefont
  {Negele}},\ }\href {\doibase 10.1103/PhysRevC.15.1359} {\bibfield  {journal}
  {\bibinfo  {journal} {Phys. Rev. C}\ }\textbf {\bibinfo {volume} {15}},\
  \bibinfo {pages} {1359} (\bibinfo {year} {1977})}\BibitemShut {NoStop}%
\bibitem [{\citenamefont {B{\l}ocki}\ and\ \citenamefont
  {Flocard}(1979)}]{blo79}%
  \BibitemOpen
  \bibfield  {author} {\bibinfo {author} {\bibfnamefont {J.}~\bibnamefont
  {B{\l}ocki}}\ and\ \bibinfo {author} {\bibfnamefont {H.}~\bibnamefont
  {Flocard}},\ }\href@noop {} {\bibfield  {journal} {\bibinfo  {journal} {Phys.
  Lett. B}\ }\textbf {\bibinfo {volume} {85}},\ \bibinfo {pages} {163}
  (\bibinfo {year} {1979})}\BibitemShut {NoStop}%
\bibitem [{\citenamefont {Dasgupta}\ \emph {et~al.}(1998)\citenamefont
  {Dasgupta}, \citenamefont {Hinde}, \citenamefont {Rowley},\ and\
  \citenamefont {Stefanini}}]{das98}%
  \BibitemOpen
  \bibfield  {author} {\bibinfo {author} {\bibfnamefont {M.}~\bibnamefont
  {Dasgupta}}, \bibinfo {author} {\bibfnamefont {D.~J.}\ \bibnamefont {Hinde}},
  \bibinfo {author} {\bibfnamefont {N.}~\bibnamefont {Rowley}}, \ and\ \bibinfo
  {author} {\bibfnamefont {A.~M.}\ \bibnamefont {Stefanini}},\ }\href@noop {}
  {\bibfield  {journal} {\bibinfo  {journal} {Ann. Rev. Nucl. Part. Sci.}\
  }\textbf {\bibinfo {volume} {48}},\ \bibinfo {pages} {401} (\bibinfo {year}
  {1998})}\BibitemShut {NoStop}%
\bibitem [{\citenamefont {Simenel}\ \emph
  {et~al.}(2013{\natexlab{a}})\citenamefont {Simenel}, \citenamefont {Keser},
  \citenamefont {Umar},\ and\ \citenamefont {Oberacker}}]{sim13b}%
  \BibitemOpen
  \bibfield  {author} {\bibinfo {author} {\bibfnamefont {C.}~\bibnamefont
  {Simenel}}, \bibinfo {author} {\bibfnamefont {R.}~\bibnamefont {Keser}},
  \bibinfo {author} {\bibfnamefont {A.~S.}\ \bibnamefont {Umar}}, \ and\
  \bibinfo {author} {\bibfnamefont {V.~E.}\ \bibnamefont {Oberacker}},\ }\href
  {\doibase 10.1103/PhysRevC.88.024617} {\bibfield  {journal} {\bibinfo
  {journal} {Phys. Rev. C}\ }\textbf {\bibinfo {volume} {88}},\ \bibinfo
  {pages} {024617} (\bibinfo {year} {2013}{\natexlab{a}})}\BibitemShut
  {NoStop}%
\bibitem [{\citenamefont {Simenel}\ \emph
  {et~al.}(2013{\natexlab{b}})\citenamefont {Simenel}, \citenamefont
  {Dasgupta}, \citenamefont {Hinde},\ and\ \citenamefont {Williams}}]{sim13c}%
  \BibitemOpen
  \bibfield  {author} {\bibinfo {author} {\bibfnamefont {C.}~\bibnamefont
  {Simenel}}, \bibinfo {author} {\bibfnamefont {M.}~\bibnamefont {Dasgupta}},
  \bibinfo {author} {\bibfnamefont {D.~J.}\ \bibnamefont {Hinde}}, \ and\
  \bibinfo {author} {\bibfnamefont {E.}~\bibnamefont {Williams}},\ }\href
  {\doibase 10.1103/PhysRevC.88.064604} {\bibfield  {journal} {\bibinfo
  {journal} {Phys. Rev. C}\ }\textbf {\bibinfo {volume} {88}},\ \bibinfo
  {pages} {064604} (\bibinfo {year} {2013}{\natexlab{b}})}\BibitemShut
  {NoStop}%
\bibitem [{\citenamefont {Simenel}\ \emph {et~al.}(2001)\citenamefont
  {Simenel}, \citenamefont {Chomaz},\ and\ \citenamefont {de~France}}]{sim01}%
  \BibitemOpen
  \bibfield  {author} {\bibinfo {author} {\bibfnamefont {C.}~\bibnamefont
  {Simenel}}, \bibinfo {author} {\bibfnamefont {P.}~\bibnamefont {Chomaz}}, \
  and\ \bibinfo {author} {\bibfnamefont {G.}~\bibnamefont {de~France}},\ }\href
  {\doibase 10.1103/PhysRevLett.86.2971} {\bibfield  {journal} {\bibinfo
  {journal} {Phys. Rev. Lett.}\ }\textbf {\bibinfo {volume} {86}},\ \bibinfo
  {pages} {2971} (\bibinfo {year} {2001})}\BibitemShut {NoStop}%
\bibitem [{\citenamefont {Simenel}\ and\ \citenamefont {Chomaz}(2003)}]{sim03}%
  \BibitemOpen
  \bibfield  {author} {\bibinfo {author} {\bibfnamefont {C.}~\bibnamefont
  {Simenel}}\ and\ \bibinfo {author} {\bibfnamefont {P.}~\bibnamefont
  {Chomaz}},\ }\href {\doibase 10.1103/PhysRevC.68.024302} {\bibfield
  {journal} {\bibinfo  {journal} {Phys. Rev. C}\ }\textbf {\bibinfo {volume}
  {68}},\ \bibinfo {pages} {024302} (\bibinfo {year} {2003})}\BibitemShut
  {NoStop}%
\bibitem [{\citenamefont {Simenel}\ and\ \citenamefont {Chomaz}(2009)}]{sim09}%
  \BibitemOpen
  \bibfield  {author} {\bibinfo {author} {\bibfnamefont {C.}~\bibnamefont
  {Simenel}}\ and\ \bibinfo {author} {\bibfnamefont {P.}~\bibnamefont
  {Chomaz}},\ }\href {\doibase 10.1103/PhysRevC.80.064309} {\bibfield
  {journal} {\bibinfo  {journal} {Phys. Rev. C}\ }\textbf {\bibinfo {volume}
  {80}},\ \bibinfo {pages} {064309} (\bibinfo {year} {2009})}\BibitemShut
  {NoStop}%
\bibitem [{\citenamefont {Maruhn}\ \emph {et~al.}()\citenamefont {Maruhn},
  \citenamefont {Reinhard}, \citenamefont {Stevenson},\ and\ \citenamefont
  {Umar}}]{mar14}%
  \BibitemOpen
  \bibfield  {author} {\bibinfo {author} {\bibfnamefont {J.~A.}\ \bibnamefont
  {Maruhn}}, \bibinfo {author} {\bibfnamefont {P.-G.}\ \bibnamefont
  {Reinhard}}, \bibinfo {author} {\bibfnamefont {P.~D.}\ \bibnamefont
  {Stevenson}}, \ and\ \bibinfo {author} {\bibfnamefont {A.~S.}\ \bibnamefont
  {Umar}},\ }\href@noop {} {\enquote {\bibinfo {title} {The tdhf code sky3d},}\
  }\bibinfo {note} {ArXiv:1310.5946}\BibitemShut {NoStop}%
\bibitem [{\citenamefont {Simenel}(2010)}]{sim10b}%
  \BibitemOpen
  \bibfield  {author} {\bibinfo {author} {\bibfnamefont {C.}~\bibnamefont
  {Simenel}},\ }\href {\doibase 10.1103/PhysRevLett.105.192701} {\bibfield
  {journal} {\bibinfo  {journal} {Phys. Rev. Lett.}\ }\textbf {\bibinfo
  {volume} {105}},\ \bibinfo {pages} {192701} (\bibinfo {year}
  {2010})}\BibitemShut {NoStop}%
\bibitem [{\citenamefont {Simenel}(2011)}]{sim11}%
  \BibitemOpen
  \bibfield  {author} {\bibinfo {author} {\bibfnamefont {C.}~\bibnamefont
  {Simenel}},\ }\href {\doibase 10.1103/PhysRevLett.106.112502} {\bibfield
  {journal} {\bibinfo  {journal} {Phys. Rev. Lett.}\ }\textbf {\bibinfo
  {volume} {106}},\ \bibinfo {pages} {112502} (\bibinfo {year}
  {2011})}\BibitemShut {NoStop}%
\bibitem [{\citenamefont {Schmidt}\ \emph {et~al.}(2000)\citenamefont
  {Schmidt}, \citenamefont {Steinh\"{a}user}, \citenamefont {B\"{o}kstiegel},
  \citenamefont {Grewe}, \citenamefont {Heinz}, \citenamefont {Junghans},
  \citenamefont {Benlliure}, \citenamefont {Clerc}, \citenamefont {de~Jong},
  \citenamefont {M\"{u}ller}, \citenamefont {Pf\"{u}tzner},\ and\ \citenamefont
  {Voss}}]{sch00}%
  \BibitemOpen
  \bibfield  {author} {\bibinfo {author} {\bibfnamefont {K.-H.}\ \bibnamefont
  {Schmidt}}, \bibinfo {author} {\bibfnamefont {S.}~\bibnamefont
  {Steinh\"{a}user}}, \bibinfo {author} {\bibfnamefont {C.}~\bibnamefont
  {B\"{o}kstiegel}}, \bibinfo {author} {\bibfnamefont {A.}~\bibnamefont
  {Grewe}}, \bibinfo {author} {\bibfnamefont {A.}~\bibnamefont {Heinz}},
  \bibinfo {author} {\bibfnamefont {A.}~\bibnamefont {Junghans}}, \bibinfo
  {author} {\bibfnamefont {J.}~\bibnamefont {Benlliure}}, \bibinfo {author}
  {\bibfnamefont {H.-G.}\ \bibnamefont {Clerc}}, \bibinfo {author}
  {\bibfnamefont {M.}~\bibnamefont {de~Jong}}, \bibinfo {author} {\bibfnamefont
  {J.}~\bibnamefont {M\"{u}ller}}, \bibinfo {author} {\bibfnamefont
  {M.}~\bibnamefont {Pf\"{u}tzner}}, \ and\ \bibinfo {author} {\bibfnamefont
  {B.}~\bibnamefont {Voss}},\ }\href {\doibase
  http://dx.doi.org/10.1016/S0375-9474(99)00384-X} {\bibfield  {journal}
  {\bibinfo  {journal} {Nucl. Phys. A}\ }\textbf {\bibinfo {volume} {665}},\
  \bibinfo {pages} {221 } (\bibinfo {year} {2000})}\BibitemShut {NoStop}%
\end{thebibliography}%

\end{document}